\documentclass[12pt,english,american]{article}
\usepackage[latin9]{inputenc}
\usepackage{geometry}
\usepackage{babel}
\usepackage{endnotes}
\usepackage{amsmath}
\usepackage{amssymb}
\usepackage{fixltx2e}
\usepackage{esint}
\usepackage[unicode=true]
 {hyperref}

\makeatletter

\DeclareRobustCommand{\greektext}{%
  \fontencoding{LGR}\selectfont\def\encodingdefault{LGR}}
\DeclareRobustCommand{\textgreek}[1]{\leavevmode{\greektext #1}}
\DeclareFontEncoding{LGR}{}{}
\DeclareTextSymbol{\~}{LGR}{126}
\DeclareRobustCommand{\cyrtext}{%
  \fontencoding{T2A}\selectfont\def\encodingdefault{T2A}}
\DeclareRobustCommand{\textcyr}[1]{\leavevmode{\cyrtext #1}}
\AtBeginDocument{\DeclareFontEncoding{T2A}{}{}}

 \let\footnote=\endnote

\makeatother

\begin{document}

\title{MODIFIED KLEIN-GORDON-FOCK EQUATIONS BASED ON ONE-DIMENSIONAL CHAOTIC
DYNAMICS AND GROUPS WITH BROKEN SYMMETRY}

%\author{D.B. Volov}

\author{D.B.\,Volov %
\thanks{Russia, Samara State University of Transport, e-mail: volovdm@mail.ru,
www.volovdm2.narod2.ru%
} \\
}

\maketitle
%Russia, Samara State University of Transport, e-mail: volovdm@mail.ru,
%www.volovdm2.narod2.ru
\begin{abstract}
Modified Klein-Gordon-Fock equations were obtained on the basis of
one-dimensional chaotic dynamics. The original Lagrangians were found.
The concepts of \textit{m}-exponential map and groups with broken
symmetry are introduced. A system of bitrial orthogonal functions
is considered.
\end{abstract}

\section*{Introduction}

The study of one-dimensional point maps in the form of \foreignlanguage{english}{$x_{n+1}\rightarrow f(x_{n})$}
\cite{VolovDyn} has revealed that some dynamics exhibit specific
behavior.

It has been shown that unlike the Verhulst-Pearl one-dimensional map
\cite{Verhulst} \foreignlanguage{english}{$x_{n+1}\rightarrow qx_{n}(1-x_{n})$},
and the Ricker discrete model \cite{Ricker} \foreignlanguage{english}{$x_{n+1}\rightarrow qx_{n}\exp(-x_{n})$},
\foreignlanguage{english}{$q\in R$}, the bifurcation diagram of the
generalized Verhulst-Ricker-Planck dynamics (VRP) \cite{VolovDyn,VolovArXiv}:
\foreignlanguage{english}{
\[
x_{n+1}\rightarrow\frac{qx_{n}^{\Phi}}{\exp(x_{n})+\alpha},\;\alpha\in R,\;\Phi\in R
\]
}along with characteristic cascades of period doubling bifurcations,
periodic windows, etc. exhibits a number of new properties: 

1. With \foreignlanguage{english}{$\Phi=1$} and $\alpha$ approaching
$-1$ from the right, the VRP chaotic component is weakening, whereas
with $\alpha=-1$ it disappears. Thus, the system \textquotedblleft{}is
cleaning\textquotedblright{} to remove chaotic bifurcations and leaving
one and only one bifurcation. 

2. With \foreignlanguage{english}{$\Phi=-2$} and $\alpha$ approaching
$1/137$ from the left, in the VRP dynamics

\selectlanguage{english}%
\begin{equation}
r_{n+1}\rightarrow\frac{2q_{1}\mu}{r_{n}^{2}(e^{\mu r_{n}}+\alpha)},\label{eq:volovdm:fourrats}
\end{equation}
\foreignlanguage{american}{where }$r,\: q_{1},\:\mu\in R$\foreignlanguage{american}{,
the chaotic component is weakening, and a new characteristic picture
of limited bifurcations occurs \textendash{} \textquotedblleft{}four
rats\textquotedblright{} \cite{VolovArXiv}.}

\selectlanguage{american}%
Thus, there exist two limit values of the dimensionless parameter:\foreignlanguage{english}{
$\alpha=-1$} and \foreignlanguage{english}{$\alpha\approx+1/137$}.
When the bifurcation diagram is approaching the first value it \textquotedblleft{}is
degenerating into two branches\textquotedblright{}; when the bifurcation
diagram is approaching the second value it has the definite number
of bifurcations on the entire area of existence with the exception
of the strictly limited zone with the chaotic component. Let us combine
the two extreme cases and write the VRP limit map as:

\selectlanguage{english}%
\begin{equation}
(\mu x)_{n+1}\rightarrow\frac{-2q_{1}\mu_{n}^{3}}{(\mu x)_{n}^{2}(e^{(\mu x)_{n}}+\alpha)}.\label{eq:volovdm:limmap}
\end{equation}

\selectlanguage{american}%
When \foreignlanguage{english}{$x$} is constant and \foreignlanguage{english}{$\alpha=-1$}
we get the first limit dynamics; when $\mu$ is constant and \foreignlanguage{english}{$\alpha\approx+1/137$},
the second limit dynamics is obtained. There have been attempts to
use map properties (\foreignlanguage{english}{\ref{eq:volovdm:limmap}})
when modeling atom nuclei with the Woods-Saxon potential \cite{Trunev}
and modeling fields of different nature \cite{VolovBitr}.

\section{Modified Klein-Gordon-Fock equations}

The question is bound to arise: if the $\alpha$ parameter is included
into the exponent according to the scheme \foreignlanguage{english}{$\exp(-x)\rightarrow\frac{1}{\exp(x)+\alpha}$},
what will be changes in the differential equations of motion due to
this substitution? As an example let us consider the operation of
substitution in the Yukawa potential \foreignlanguage{english}{
\[
\varphi=-\frac{const\cdot e^{-\mu r}}{r}.
\]
}This potential is known as a stationary solution of the equation
similar to the Klein-Gordon equation. It approximates how the residual
forces of strong interaction fields behave in the stationary spherically
symmetric case \cite{Bogoluybov}.

It is shown that the equation solved in the axially symmetric stationary
case as the modified Yukawa potential 

\selectlanguage{english}%
\begin{equation}
\varphi=-\frac{const}{r(e^{\mu r}+\alpha)},\label{eq:volovdm:mYukawa}
\end{equation}
\foreignlanguage{american}{and a partial wave solution }

\begin{equation}
\psi=\frac{\psi_{0}}{e^{i\mu z}+\alpha},\label{eq:volovdm:wawemYukawa}
\end{equation}
\foreignlanguage{american}{corresponds to the system of linear differential
equations of second order: }

\begin{equation}
((1+\alpha e^{-\mu x})\square+2\mu^{k}\partial_{k}+\mu^{2})\varphi_{i}=0,\label{eq:volovdm:mKGF}
\end{equation}
\foreignlanguage{american}{where in the exponent }$\mu x$\foreignlanguage{american}{
is understood as the dot product of 4-vectors with the components
}$\mu_{k}$\foreignlanguage{american}{ and }$x^{k}$\foreignlanguage{american}{,
Einstein summation convention is used, $\square$ is the wave operator,
}$\varphi_{i}$\foreignlanguage{american}{ is the multicomponent field,
}$z$\foreignlanguage{american}{ in (}\ref{eq:volovdm:wawemYukawa}\foreignlanguage{american}{)
is the preferential direction of wave propagation.}

\selectlanguage{american}%
The equations (\foreignlanguage{english}{\ref{eq:volovdm:mKGF}})
are restored according to the known solution. To build (\foreignlanguage{english}{\ref{eq:volovdm:mKGF}})
we have twice differentiated the expression (\foreignlanguage{english}{\ref{eq:volovdm:wawemYukawa}})
with respect to \foreignlanguage{english}{$z$}; the first and second
derivatives formed the one-dimensional equation with the variable
\foreignlanguage{english}{$z$} to be solved as (\foreignlanguage{english}{\ref{eq:volovdm:wawemYukawa}}):

\selectlanguage{english}%
\[
((1+\alpha e^{-i\mu z})\partial_{z}^{2}+2i\mu\cdot\partial_{z}-\mu^{2})\psi=0.
\]

\selectlanguage{american}%
Generalization of the last equation for any 4-vector \foreignlanguage{english}{$\mu_{k}$}
(as the wave eigenvector) and the multicomponent field \foreignlanguage{english}{$\varphi_{i}$}
resulted in (\foreignlanguage{english}{\ref{eq:volovdm:mKGF}}). Here
\foreignlanguage{english}{$\mu^{2}=(mc/\hbar)^{2}$}, \foreignlanguage{english}{$c$}
is the speed of light, \foreignlanguage{english}{$\hbar$} is the
reduced Planck constant, \foreignlanguage{english}{$m$} is the mass.

The wave solution is obtained from (\foreignlanguage{english}{\ref{eq:volovdm:mKGF}})
when \foreignlanguage{english}{$\mu\rightarrow i\mu$}. As the solutions
(\foreignlanguage{english}{\ref{eq:volovdm:mKGF}}) with \foreignlanguage{english}{$\alpha=0$}
is a part of the solutions of the Klein-Gordon-Fock (KGF) equation
with the positive exponent, we call the equations (\foreignlanguage{english}{\ref{eq:volovdm:mKGF}})
modified Klein-Gordon-Fock equations (\textit{m}-\textcyr{\char202}GF).
The second conjugate part of the KGF solutions, with the negative
exponent, corresponds to the \textit{m}-\textcyr{\char202}GF solutions
when \foreignlanguage{english}{$\mu\rightarrow-\mu$}.

\textbf{Note:} to meet the conditions of (\foreignlanguage{english}{\ref{eq:volovdm:mYukawa}}),
the stationary \textit{m}-\textcyr{\char202}GF equation in spherical
coordinates requires the presence of the field source in its right-hand
side: 

\selectlanguage{english}%
\[
(1+\alpha e^{-\mu r})\frac{\partial^{2}(r\varphi)}{r\partial r^{2}}+2\mu\frac{\partial(r^{2}\varphi)}{r^{2}\partial r}+\mu^{2}\varphi=\frac{2q_{1}\mu}{r^{2}(e^{\mu r}+\alpha)}.
\]
\foreignlanguage{american}{which, however strange it may seem, appears
in the VRP limit map in the formula (}\ref{eq:volovdm:fourrats}\foreignlanguage{american}{).}

\selectlanguage{american}%
What we are dealing with is the restoration of the \textit{m}-\textcyr{\char202}GF
equations themselves (\foreignlanguage{english}{\ref{eq:volovdm:mKGF}})
provided the known partial solution is available which is of special
interest to us due to specific behavior of chaotic dynamics. \textit{m}-\textcyr{\char202}GF
equations expand the class of KGF equations with adding the nondimensional
parameter \textgreek{a}, destroying the symmetry of the exponential
map.

\section{Modified equations Lagrangian}

In \cite{VolovBitr} the focus was on the properties of the function
\foreignlanguage{english}{$m\exp(\mu x)=\frac{const}{e^{-\mu x}+\alpha}$},
\foreignlanguage{english}{$\mu\in C$}, \foreignlanguage{english}{$\mu\in H$},
which we call the bitrial exponent. The term \textquotedblleft{}bitrial\textquotedblright{}
(from \textquotedblleft{}bi\textquotedblright{} - 2 to \textquotedblleft{}tri\textquotedblright{},
which in Russian means 3) was introduced in the earlier works on formal
logic as the transition from classic logic (with the proposition values
\textquotedblleft{}yes\textquotedblright{} (1), \textquotedblleft{}no\textquotedblright{}
(0)) to expanded logic based on the three crisp values: \textquotedblleft{}yes\textquotedblright{}
(1), \textquotedblleft{}no\textquotedblright{} (\textendash{}1) as
the opposite to \textquotedblleft{}yes\textquotedblright{} (1), and
\textquotedblleft{}absence\textquotedblright{} (0). There is a correspondence
between the Verhulst dynamics and VRP similar to the above mentioned
correspondence.

As the bitrial exponent, unlike the ordinary one, is limited on the
entire range of definition, it was hypothesized that substituting
the exponents in the corresponding functional dependences with the
bitrial exponent \foreignlanguage{english}{$e^{\mu x}\rightarrow\frac{1}{e^{-\mu x}+\alpha}$
}and the subsequent approaching $\alpha$ to \foreignlanguage{english}{$0$},
we can develop new methods of regularization to solve the problems
where the Fourier transform, the bilateral Laplace transform, other
integral transformations, gauge transformations for the field functions
with divergent propagators, Kruskal coordinates et al. are used.

Having found (\foreignlanguage{english}{\ref{eq:volovdm:mKGF}}),
let us restore the Lagrangians from which these equations are obtained.

It turned out that even in the one-dimensional case there does not
exist any Lagrangian for equations (\foreignlanguage{english}{\ref{eq:volovdm:mKGF}}).
You may refer to the special theorem \cite{VolovBitr}. On the other
hand, there does exist the Lagrangian \foreignlanguage{english}{
\[
\mathcal{L}=(\alpha+e^{\pm\mu x})\partial_{n}U_{i}\,\partial^{n}U_{i}^{*}-\mu^{2}U_{i}U_{i}^{*}e^{\pm\mu x},
\]
}similar to the motion equation (\foreignlanguage{english}{\ref{eq:volovdm:mKGF}}) 

\selectlanguage{english}%
\[
(1+\alpha e^{\mp\mu x})\square U_{i}\pm\mu^{n}\partial_{n}U_{i}+\mu^{2}U_{i}=0.
\]
\foreignlanguage{american}{The absence of the coefficient \textquotedblleft{}2\textquotedblright{}
before the second member of the last equation is critical for building
a Lagrangian. To solve this problem we write (}\ref{eq:volovdm:mKGF}\foreignlanguage{american}{)
simultaneously with the source on the right}
\[
(1+\alpha e^{\mp\mu x})\square G_{i}\pm\mu^{n}\partial_{n}G_{i}+\mu^{2}G_{i}=\mp\mu^{n}\partial_{n}G_{i}
\]
\foreignlanguage{american}{bearing in mind that the Lagrangian exists
only for the left side of the equation. Here }$U,\: G$\foreignlanguage{american}{
are some complex fields.}

\selectlanguage{american}%
There is a reason to assume that the \textit{m}-\textcyr{\char202}GF
equation describes the self-acting system \textquotedblleft{}field
- field source\textquotedblright{} without separating the field source
and the free field as it is done in perturbation theory, in quantum
field theory. Then, the propagator \foreignlanguage{english}{$G$}
to the \textit{m}-\textcyr{\char202}GF equation is the exponential
eigenfunction of the homogenous equation (\foreignlanguage{english}{\ref{eq:volovdm:mKGF}}).

Let us agree to use the symbol \foreignlanguage{english}{«\textit{m}»}
in the function names where the substitution took place according
to the scheme \foreignlanguage{english}{$e^{\theta}\rightarrow\frac{1}{e^{-\theta}+\alpha}$},
representing \foreignlanguage{english}{$\forall f(x),\; mf(x)\equiv f(\frac{1}{x^{-1}+\alpha})$}.
For example, let us represent the function \foreignlanguage{english}{$\frac{1}{e^{-i\theta}+\alpha}$}
(where $\theta$ is the number, the angle) as \foreignlanguage{english}{$me^{i\theta}$}or
as \foreignlanguage{english}{$m\exp(i\theta)$}. Such substitution
in \cite{VolovBitr} we call bitrial. Hereafter the symbol \foreignlanguage{english}{«\textit{m}»}
will be put before other objects and mathematical definitions modified
according to this scheme.

\section{Groups with broken symmetry}

Further, it is appropriate to introduce the notion of groups with
broken symmetry \foreignlanguage{english}{$G(\alpha)$} and single
out the class of special unitary groups \foreignlanguage{english}{$SU(n,\,\alpha)$}.
Let us begin to build such groups with the building of a new group
algebra and considering the group $U(1)$.

Complex numbers \foreignlanguage{english}{$g$} and \foreignlanguage{english}{$g_{1}$},
equal to unity in modules, can be represented either as \foreignlanguage{english}{$g=e^{i\theta}$},
\foreignlanguage{english}{$g_{1}=e^{i\theta_{1}}$} or \foreignlanguage{english}{$g=\frac{1}{e^{-i\theta^{\prime}}+\alpha}$},
\foreignlanguage{english}{$g_{1}=\frac{1}{e^{-i\theta_{1}^{\prime}}+\alpha}$};
hence the product of \foreignlanguage{english}{$g$} and \foreignlanguage{english}{$g_{1}$}
gives the third element \foreignlanguage{english}{$g_{2}$} of the
same group. However, this third element is the element of the group
$U(1)$. In real number algebra the element \foreignlanguage{english}{$g_{2}$}
of the group $U(1)$ is represented as \foreignlanguage{english}{$gg_{1}=g_{2}=e^{i(\theta+\theta_{1})}$},
whereas in new algebra this very element of the group $U(1)$ is defined
as \foreignlanguage{english}{$gg_{1}=g_{2}=me^{i(\theta^{\prime}\oplus\theta_{1}^{\prime})}$},
where by $\oplus$ we mean the addition operation resulting in the
identity: 

\selectlanguage{english}%
\begin{gather}
gg_{1}=me^{i\theta^{\prime}}me^{i\theta_{1}^{\prime}}=me^{i(\theta^{\prime}\oplus\theta_{1}^{\prime})}=\label{eq:volovdm:msumma}\\
=\left(\frac{1}{e^{-i\theta^{\prime}}+\alpha}\right)\left(\frac{1}{e^{-i\theta_{1}^{\prime}}+\alpha}\right)=\nonumber \\
=\frac{1}{e^{-i(\theta^{\prime}+\theta_{1}^{\prime})}+\alpha(e^{-i\theta^{\prime}}+e^{-i\theta_{1}^{\prime}})+\alpha^{2}}.\nonumber 
\end{gather}

\selectlanguage{american}%
So, we introduce the new algebra \foreignlanguage{english}{$\{\boldsymbol{\Theta^{\prime}},\oplus\}$}
in the parameter space \foreignlanguage{english}{$\boldsymbol{\Theta^{\prime}}\equiv\{\theta^{\prime}\}\subset R$},
and define within constant value the binary operation \foreignlanguage{english}{(\ref{eq:volovdm:msumma})},
satisfying the following axioms:

1) in the group \foreignlanguage{english}{$U(1)$} there exists a
unit element \foreignlanguage{english}{$g_{0}$}, for which \foreignlanguage{english}{
\[
\theta_{0}^{\prime}=i\ln(1-\alpha);
\]
}

2) for each \foreignlanguage{english}{$g$} element of the group $U(1)$
there exists an inverse element \foreignlanguage{english}{$g^{-1}$}
with the relation 

\selectlanguage{english}%
\[
(\theta^{\prime})^{-1}=i\ln\left(\frac{1-\alpha e^{-\theta^{\prime}}-\alpha^{2}}{e^{-\theta^{\prime}}+\alpha}\right);
\]

\selectlanguage{american}%
3) the multiplication operation of the elements of the group $U(1)$
is associative:

\selectlanguage{english}%
\[
g(g_{1}g_{2})=(gg_{1})g_{2}.
\]

\selectlanguage{american}%
To prove the commutative, associative and distributive properties
in such map let us have a look at the function graph \foreignlanguage{english}{$\varphi^{\prime}=\frac{1}{e^{-i\theta^{\prime}}+\alpha}$}
in the complex plane. It is a circle with the radius \foreignlanguage{english}{$\frac{1}{1-\alpha^{2}}$}
and the center in the point \foreignlanguage{english}{$\frac{-\alpha}{1-\alpha^{2}}$}.
Therefore to provide a more accurate comparison, we normalize the
function $\varphi^{\prime}$: \foreignlanguage{english}{$\varphi^{\prime}=\frac{1-\alpha^{2}}{e^{-i\theta^{\prime}}+\alpha}$}.

At first, on the section $[0,\:\frac{\pi}{2}]$ the rate of change
$\varphi^{\prime}$ lags behind the uniform rate of rotation \foreignlanguage{english}{$\varphi_{R}=(e^{-i\theta}-\alpha)$}
with the center in $-\alpha$. Here, by uniform rotation we mean sequential
addition operation by one and the same amount of the angle change
at every step. On \foreignlanguage{english}{$[\frac{\pi}{2},\:\pi]$}
the function $\varphi^{\prime}$ catches up, so at the point $\theta$=$\pi$
$\varphi^{\prime}$ arrives at the same time as \foreignlanguage{english}{$\varphi_{R}$}.
As the parameters $\theta^{\prime}$ and $\theta$ change continuously
and monotonously, within the range $[0,\,\pi]$ there is always the
$\theta$, which unambiguously corresponds to the $\theta^{\prime}$
from the range $[0,\,\pi]$. Then, on the $[0,\,\pi]$ the bitrial
exponent $\varphi^{\prime}$ overtakes the uniform circular motion
$\varphi_{R}$, but on the remaining section lags behind it; so, at
$\theta=2\pi$ both arrive at the same time. As the parameters $\theta^{\prime}$
and $\theta$ change continuously and monotonously, within the range
$[\pi,\,2\pi]$ there is always the $\theta$, which unambiguously
corresponds to the $\theta^{\prime}$ from the range $[\pi,\,2\pi]$.
Thus, the rotation around $\varphi^{\prime}$ is non-uniform. On average,
on the section $[0,\,2\pi]$, the result will be the same in both
cases. Let us distinguish between them within $[0,2\pi]$ as exponential
and \textit{m}-exponential maps.

Between the angle $\theta$ measured in the coordinate system\foreignlanguage{english}{$\varphi_{R}$}
and the angle $\theta^{\prime}$ measured in the coordinate system
$\varphi^{\prime}$, on the section $[0,\,2\pi]$ there exists a uniform
correspondence: 

\selectlanguage{english}%
\begin{equation}
\theta=i\ln\left(\frac{1+\alpha e^{i\theta^{\prime}}}{e^{i\theta^{\prime}}+\alpha}\right).\label{eq:volovdm:isomorphism}
\end{equation}

\selectlanguage{american}%
In addition to the center of the circle translation, in the general
case we are to consider the rotation of the group center including
the phase \foreignlanguage{english}{$\theta_{W}$} in (\foreignlanguage{english}{\ref{eq:volovdm:isomorphism}}).
It may be important for gauge transformations of the groups with broken
symmetry which are beyond the scope of the present paper. 

Defining the addition operation \foreignlanguage{english}{$\theta^{\prime}\oplus\theta_{1}^{\prime}$}
in the \textit{m}-exponential map, we construct the group algebra,
in this case for $U(1)$, different from the real number algebra with
the operation \foreignlanguage{english}{$\theta+\theta_{1}$} in the
exponential map \foreignlanguage{english}{$e^{i\theta}$} of the \foreignlanguage{english}{$g$}
group elements, in our case complex numbers equal to unity in modules.
Thus, for the elements of one and the same group $U(1)$ two different
algebras are constructed with the different binary operation; their
maps being different.

Everything said above on the \textit{m}-map for the group $U(1)$,
can be generalized with regard to any group, allowing the exponential
map \foreignlanguage{english}{$e^{\sum\limits _{i}T_{i}\theta_{i}}$}.
The following theorem can be formulated:

T h e\ \ \  t h e o r e m. Any group \foreignlanguage{english}{$G$}
featuring the exponential map \foreignlanguage{english}{$e^{\sum\limits _{i}T_{i}\theta_{i}}$}
with the generators \foreignlanguage{english}{$T_{i}$} and parameters
\foreignlanguage{english}{$\boldsymbol{\Theta}\equiv\{\theta\}$},
allows an isomorphic \textit{m}-exponential map \foreignlanguage{english}{$\left[e^{\sum\limits _{i}T_{i}\theta_{i}^{\prime}}+\alpha\right]^{-1}$}
with the same generators \foreignlanguage{english}{$T_{i}$} and \textit{m}-algebra
\foreignlanguage{english}{$\{\boldsymbol{\Theta^{\prime}},\oplus\}$}
over the parameters \foreignlanguage{english}{$\boldsymbol{\Theta^{\prime}}$},
where $\alpha$ is of the same rank as \foreignlanguage{english}{$G$},
\foreignlanguage{english}{$|\alpha|>0$}. 

Having briefly described the \textit{m}-algebra, let us introduce
the notion of a group with broken symmetry with the source group $U(1)$
serving as an example. As the $U(1,\,\alpha)$ group with broken symmetry
we take the multiplicative Abel group of all complex numbers, equal
to unity in modulus, where each element is obtained by applying the
sequential addition operation to the infinitesimal value of the angle
\foreignlanguage{english}{$\Delta\theta^{\prime}\:$} in the \textit{m}-algebra.
In other words, the group $U(1,\,\alpha)$ refers to the group obtained
by the uniform rotation of the \textit{m}-algebra elements at the
uniform angular velocity \foreignlanguage{english}{$\omega^{\prime}=\frac{d\theta^{\prime}}{dt}=\frac{\Delta\theta^{\prime}}{\Delta t}=const$}.
In the complex plane we observe the non-uniform rotation of the group
elements. Each element \foreignlanguage{english}{$g=\frac{1}{e^{-i\theta^{\prime}}+\alpha}$}
of the group $U(1,\,\alpha)$ is the element \foreignlanguage{english}{$g=e^{i\theta}$}
of the group $U(1)$; between the elements of these groups there exists
an unambiguous correspondence:

\selectlanguage{english}%
\begin{gather*}
\theta=\theta^{\prime}+i\ln\left(\frac{1+\alpha e^{i\theta^{\prime}}}{1+\alpha e^{-i\theta^{\prime}}}\right),\\
\theta^{\prime}=\theta+i\ln\left(\frac{1-\alpha e^{i\theta}}{1-\alpha e^{-i\theta}}\right).
\end{gather*}

\selectlanguage{american}%
Hence, there is every reason to write the algebraic operation \foreignlanguage{english}{(\ref{eq:volovdm:msumma})}
as: 

\selectlanguage{english}%
\[
\theta_{1}^{\prime}\oplus\theta_{2}^{\prime}=\left(\theta_{1}+\theta_{2}\right)+i\ln\left(\frac{1-\alpha e^{i\left(\theta_{1}+\theta_{2}\right)}}{1-\alpha e^{-i\left(\theta_{1}+\theta_{2}\right)}}\right).
\]

\selectlanguage{american}%
Provided the rotation frequency \foreignlanguage{english}{$\omega=\frac{d\theta}{dt}$}
is high enough, the difference in the algebras and non-uniform rotation
in the groups $U(1,\,\alpha)$ and $U(1)$ are not observed. All elements
of the group $U(1,\,\alpha)$ are elements of the group $U(1)$, obtained
by translating the group center in the complex plane. $U(1,\,\alpha)$
differs from $U(1)$ because of the non-uniform rotation of the group
elements within the section \foreignlanguage{english}{$[0,\:2\pi]$}.

Thus, we get an Abel compact group $U(1,\,\alpha)$, an isomorphic
$U(1)$ with the translation and non-uniform rotation. For this group
the multiplication operation (\ref{eq:volovdm:msumma}) of the elements
\foreignlanguage{english}{$g$} is called addition of the elements
$\theta^{\prime}$; the unit element \foreignlanguage{english}{$\theta_{0}^{\prime}$}
is called null. Addition rules for this group enable us to define
such trigonometric function as the bitrial cosine, the sine 

\selectlanguage{english}%
\begin{gather*}
m\cos\theta=\alpha+\frac{1-\alpha^{2}}{2}\left(\frac{1}{e^{-i\theta}+\alpha}+\frac{1}{e^{i\theta}+\alpha}\right),\\
m\sin\theta=\frac{1-\alpha^{2}}{2i}\left(\frac{1}{e^{-i\theta}+\alpha}-\frac{1}{e^{i\theta}+\alpha}\right),
\end{gather*}
\foreignlanguage{american}{the bitrial logarithm }$m\ln\theta=\ln\left(\frac{\theta}{1+\alpha\theta}\right)$\foreignlanguage{american}{,
hyperbolic functions, etc. It should be noted that there are no changes
in the whole arsenal of known relations, provided the addition operation
$\oplus$ is determined for them.}

\selectlanguage{american}%
Further, the Kelley-Dixon recursive procedure enables us to expand
the notion of the group with broken symmetry $U(1,\,\alpha)$ from
complex numbers to quaternions (isomorphically to $SU(2,\,\alpha)$),
octonions and sedenions. From practical point of view, of special
interest are the groups with the dot product, with \foreignlanguage{english}{$\theta=\mu x$}.
For instance, \foreignlanguage{english}{$\theta=\mu_{0}x^{0}+\mu_{1}x^{1}+\mu_{2}x^{2}+\mu_{3}x^{3}=\omega t-\mathbf{k\mathbf{r}}$},
where \foreignlanguage{english}{$\mathbf{k}$} is the propagation
vector, \foreignlanguage{english}{$\mathbf{r}$} is the radius-vector.

Provided the matrices realizing the group $SU(2,\,\alpha)$ are uniformized
in the form of \textit{m} \foreignlanguage{english}{
\[
m\exp\left(i\frac{\theta^{\prime}}{2}\boldsymbol{\sigma\mathbf{n}}\right)=m\cos\frac{\theta^{\prime}}{2}+i\sigma\mathbf{n}\, m\sin\frac{\theta^{\prime}}{2},
\]
}where \foreignlanguage{english}{$\boldsymbol{\sigma=\left(\sigma_{1},\mathrm{\sigma_{2},\sigma_{3}}\right)}$}
are the Pauli matrices, \foreignlanguage{english}{$\mathbf{n}$} is
the 3-component real unit vector, then with small transformations
of the group $SU(2,\,\alpha)$ (the elements from the group \foreignlanguage{english}{$O(3)$})
\foreignlanguage{english}{$\mathbf{n}$} indicates the direction of
the rotation axis, whereas $\theta^{\prime}$ is the rotation angle.
The considerations related to $U(1,\,\alpha)$ and $SU(2,\,\alpha)$
are equally applicable to any other \foreignlanguage{english}{$SU(n,\,\alpha)$}.

\section{About a modified Fourier analysis}

In the \textit{m}-map, any field sources in the equations (\foreignlanguage{english}{\ref{eq:volovdm:mKGF}})
are decomposed using the trigonometrical polynominals \foreignlanguage{english}{$\varphi_{n}=m\exp\left(i\theta,\, n\right)=\left(\frac{1-\alpha^{2}}{e^{-i\theta}+\alpha}\right)^{n}$},
\foreignlanguage{english}{$n\in Z$}, obtained from $U(1,\,\alpha)$
as the direct product of the groups. Resulting infinite \textit{m}-Fourier
series with the \textit{m}-Fourier coefficients of any function \foreignlanguage{english}{$f(\theta)$}
on \foreignlanguage{english}{$[0,\:2\pi]$}

\selectlanguage{english}%
\begin{gather*}
a_{n}=\sqrt{\frac{1-\alpha^{2}}{2\pi}}\;\intop_{0}^{2\pi}f(\theta)\cdot m\exp^{*}\left(i\theta,\, n\right)d\theta=\\
=\sqrt{\frac{1-\alpha^{2}}{2\pi}}\;\intop_{0}^{2\pi}f(\theta)\cdot\left(\frac{1-\alpha^{2}}{e^{i\theta}+\alpha}\right)^{n}d\theta,\; n\in Z,
\end{gather*}
\foreignlanguage{american}{are equivalent to the Fourier series with
}$\alpha=0$\foreignlanguage{american}{ (\textquotedblleft{}{*}\textquotedblright{}
means the conjugation of functions).}

\selectlanguage{american}%
There are some peculiarities in representing conjugation. The point
is that with the next conjugation the bitrial functions are not orthogonal
in the proper sense of the word: for \foreignlanguage{english}{$\varphi_{n}$}
and \foreignlanguage{english}{$\varphi_{m}^{*}$} conjugate to it,
on the section \foreignlanguage{english}{$[0,\:2\pi]$} the identity
\foreignlanguage{english}{
\[
\intop_{0}^{2\pi}\varphi_{n}\varphi_{m}^{*}d\theta=\intop_{0}^{2\pi}\left(\frac{1-\alpha^{2}}{e^{-i\theta}+\alpha}\right)^{n}\left(\frac{1-\alpha^{2}}{e^{-i\theta}+\alpha}\right)^{-m}d\theta=\begin{cases}
2\pi, & m=n,\\
0, & m\neq n
\end{cases}
\]
}is satisfied indeed, with the exception of 

\selectlanguage{english}%
\[
\intop_{0}^{2\pi}\left(\frac{1-\alpha^{2}}{e^{-i\theta}+\alpha}\right)^{n}\left(\frac{1-\alpha^{2}}{e^{-i\theta}+\alpha}\right)^{-m}d\theta=\frac{2\pi\alpha}{1-\alpha^{2}},\; m=n+1.
\]
\foreignlanguage{american}{Naturally, with }$\alpha=0$\foreignlanguage{american}{
the system }$\left\{ \sqrt{\frac{1-\alpha^{2}}{2\pi}}me^{in\theta}|\, n\in Z\right\} $\foreignlanguage{american}{
becomes an ordinary complete orthonormalized Fourier system. Substituting
}$z\rightarrow e^{i\theta}$\foreignlanguage{american}{, }$a\rightarrow-\alpha$\foreignlanguage{american}{
in the Laurent series }$\sum\limits _{n\in Z}a_{n}\left(z-a\right)^{n}$\foreignlanguage{american}{,
we prove that with }$\alpha\neq0$\foreignlanguage{american}{ the
map is complete.}

\selectlanguage{american}%
Assigning \foreignlanguage{english}{$\varphi_{n}$} to the conjugate
function as \foreignlanguage{english}{$\varphi_{m}^{*}=m\exp\left(-i\theta,\, m\right)=\left(\frac{1-\alpha^{2}}{e^{i\theta}+\alpha}\right)^{m}$},
\foreignlanguage{english}{$m\in Z$} and evaluating the Poisson integral,
we can easily show that the system like this one forms the basis,
without any exception for \foreignlanguage{english}{$m\neq n$}.

\section*{Conclusions}

Chaotic dynamics exhibiting specific behavior allowed us to establish
the form of original equations of motion solved with these dynamics.
Obtained equations of motion were used to restore the Lagrangians.
The notions of an \textit{m}-algebra and a group with broken symmetry
are introduced. It is shown that the field sources are decomposed
into the \textit{m}-Fourier series, which is the generalization of
the Fourier series for the system of broken symmetry. The research
conducted may find its use in the quantum field theory.

\end{document}